\documentclass[a4paper,11pt]{article}
\usepackage{pos}

\title{Di-hadron Fragmentation Functions beyond LO}

\author*[a,b]{Luca Polano}
\author[a,b]{Alessandro Bacchetta}
\author[b]{Marco Radici}

\affiliation[a]{Dipartimento di Fisica ``A. Volta'', Universit\`a di Pavia,\\
Via Bassi 6, I-27100 Pavia, Italy}

\affiliation[b]{INFN - Sezione di Pavia,\\
Via Bassi 6, I-27100 Pavia, Italy}

\emailAdd{luca.polano01@universitadipavia.it}
\emailAdd{alessandro.bacchetta@unipv.it}
\emailAdd{marco.radici@pv.infn.it}

\abstract{In this talk, we 
discuss 
the extraction of Dihadron Fragmentation Functions (DiFFs) beyond 
leading order. In particular, we review the latest extraction by the MAP Collaboration of the Unpolarized Dihadron Fragmentation Functions (DiFFs) from a fit of the 2017 BELLE data 
for the inclusive $\pi^+\pi^-$ pair production 
within the same jet
in $e^+e^-$ annihilations. 
This new extraction improves 
the perturbative QCD accuracy by including NNLO calculations, 
relies 
on Monte Carlo generators 
only for flavor decomposition, 
and complements 
the traditional physics-informed parametrization with a Neural Network extraction. We also discuss the setup for the extraction of the transversely polarized DiFFs, where we revisit the NLO calculation for $e^+e^-$ annihilation into a transversely polarized quark-antiquark pair, obtaining a result different from the one reported in the literature.
 }

\FullConference{
}

\begin{document}
\maketitle

\section{Introduction}
Hadronisation is a fundamental non-perturbative phenomenon in Quantum Chromodynamics and is therefore described in terms of non-perturbative objects called Fragmentation Functions (FFs), which encode the probability for a parton to fragment into hadrons carrying a fraction $z$ of its momentum. 
In this document, we focus on Di-hadron Fragmentation Functions (DiFFs), describing the fragmentation of a parton into two hadrons produced within the same jet. 
The hadron pair is 
characterized by the 
total 
momentum fraction $z$ and invariant mass $M_h$. 
In the regime $M_h^2 \ll Q^2$, they represent a distinct nonperturbative object and obey the same evolution equations as single-hadron FFs ~\cite{CRB2007}. Within this framework, they provide an alternative method for extracting the transversity PDF in collinear factorization, which avoids some of the usual complications of TMD factorization. 
More importantly, this approach allows for the inclusion of proton-proton collision data alongside $e^+e^-$ and SIDIS measurements. Since the current state-of-the-art extractions of transversity based on this method are performed at LO, extending the analysis beyond this accuracy requires the determination of DiFFs beyond LO.
In this document, we describe the work presented in Ref.~\cite{JHEP2026}, where we revisit the DiFF extraction of Ref.~\cite{BCRB2012} by directly fitting the BELLE $e^+e^-$ data that have since become available. For the first time, the theoretical accuracy 
is 
extended 
up to NNLO, and both a physics-informed parametrization and a Neural Network (NN) approach have been employed. We also discuss recent developments in the polarized sector beyond the LO.

\section{Formalism}
In Single Dihadron Inclusive Annihilation (SDIA),
$e^+e^- \rightarrow (\pi^+ \pi^-)X$, 
an electron and a positron annihilate into a time-like virtual photon, which subsequently produces a quark-antiquark pair. The quark and antiquark then fragment into two opposite hemispheres. Each parton fragments into a residual jet and a pion pair, but only one of the two pairs is identified. The pair is characterized by momenta and masses $(P_1, M_1)$ and $(P_2, M_2)$.
To describe the process, we introduce the pair total momentum, $P_h = P_1 + P_2$, and the pair invariant mass, $M_h^2 = P_h^2$. Using the standard notations for the light-cone components of a 4-vector, we can define the light-cone fraction of the quark momentum carried by the pion pair $z =\frac{P_h^-}{q^-} = z_1+z_2$.

The differential unpolarized cross section in $z$, $M_h$, and $Q^2$ for the semi-inclusive production of a pion pair in the region $M_h^2 \ll Q^2$ can be written as in Ref.~\cite{BCR2013}, and generalized to all orders as:
\begin{equation}
\frac{d\sigma}{dz\, dM_h\, dQ^2}
= \frac{4\pi \alpha^2}{Q^2} \sum_{f=q, \overline{q}} e_f^2\, 
\sum_i \Bigl(C^f_i \otimes D_1^i \Bigr) (z, M_h; \alpha_S(Q^2), Q^2) \; ,
\label{eq:dsig0-NLO}
\end{equation}
where $\otimes$ denotes the usual convolution:
\begin{equation}
    \Bigl( C\otimes D \Bigr) (z) = \int_z^1 \frac{dy}{y}\,C(y)\, D\Bigl(\frac{z}{y}\Bigr).
\end{equation}
At LO $C_i^{f(0)}(z) = \delta_i^f \delta (1-z)$ and we recover the expression of Ref.~\cite{BCR2013}.

\section{Data analysis and flavor separation}
The center-of-mass energy of the $e^+e^- \rightarrow (\pi^+ \pi^-)X$ process at BELLE is $\sqrt{S} = 10.58,\text{GeV}$.
We impose kinematic cuts on both $M_h$ and $z$. The invariant-mass range is restricted to $0.30 < M_h < 1.30$, where the lower limit is set by the pion-pair production threshold $2m_\pi$, while the upper limit ensures the condition $M_h^2 \ll Q^2$. The momentum fraction is limited to $0.3 < z < 0.7$ in order to select the current fragmentation region and suppress contamination from exclusive processes. In order to avoid large mass corrections, we impose $\gamma_h = \frac{2M_h}{z\sqrt{s}} \leq 0.5$. To avoid the $K_S^0$ resonance peak, we exclude the bin $0.48~\text{GeV} < M_h < 0.50~\text{GeV}$. In total, we consider 344 data points. The systematic errors are point-by-point uncorrelated, but data are affected by a $100\%$ correlated luminosity uncertainty of $1.6 \%$.

From Eq.~\eqref{eq:dsig0-NLO}, 
the cross section 
is sensitive to the sum of the quark and antiquark fragmentation channels. 
In order to separate each flavor contribution, 
we need to supplement the BELLE data with information about the flavor structure. 
It 
is useful to define the ratios between the 
Monte Carlo (MC) 
flavor-tagged cross-section and the total one, 
\begin{equation}
R^q_{\text{MC}} (z, M_h, Q^2) = \frac{d\sigma^{q}_{\text{MC}}}{dz\, dM_h\, dQ^2} {\Bigg /} \frac{d\sigma}{dz\, dM_h\, dQ^2} \; ,
\label{eq:flav_ratios}
\end{equation}
where the flavor tagged cross-sections are defined as:
\begin{equation}
\frac{d\sigma^{q}_{\text{MC}}}{dz\, dM_h\, dQ^2} \\
= \frac{4\pi \alpha^2}{Q^2} e_q^2\, 
\sum_i  \Bigl[ \Bigl( C_i^q \otimes D_1^i \Bigr) + 
\Bigl( C_i^{\bar{q}} \otimes D_1^i \Bigr) \Bigr] \; .
\label{eq:dsig0-flavor}
\end{equation}
These quantities can be obtained from Monte Carlo simulations. By multiplying these 
ratios bin by bin with the experimental data, it is possible to construct four separate pseudodata sets, 
\begin{equation}
    \mathcal{F}^\mathbf{q}(z, M_h;Q^2) =  2\cdot  \dfrac{d\sigma^{exp}}{dzdM_hdQ^2} 
\cdot 
R^ {\mathbf{q}}_{MC}(z,M_h,Q).
\label{eq:ratio}
\end{equation}
By fitting these datasets separately, it is possible to extract each flavor contribution to DiFFs. 
In Eq.~\eqref{eq:ratio}, the factor of $2$ symmetrizes the quark and antiquark contributions. 
When summed together over all flavors, the $\mathcal{F}^\mathbf{q}$ reproduce the total cross section. 

\section{Parameterizations}
For the parametrization of the non-perturbative part of the DiFFs, we choose both a physics-informed functional form and an NN one, 
both 
at the scale $Q_0 = 1$ GeV. For the physics-informed parametrization, we took inspiration from the one proposed in Ref.~\cite{BCRB2012},  
where each (non-)resonant channel of the pair invariant-mass distribution is reconstructed as the sum of different quark contributions. Specifically, in 
Ref.~\cite{BCRB2012} 
the considered channels were  
the $\rho\rightarrow \pi^+ \pi^-$ at $M_h \sim 776$ MeV, the $\omega\rightarrow \pi^+ \pi^-$ at $M_h \sim 776$ MeV, the $\omega\rightarrow \pi^+ \pi^- \pi^0$ at $M_h \sim 500$ MeV and a ``continuum'' which is modeled as the fragmentation into an ``incoherent'' pion pair. In Ref.~\cite{JHEP2026}, 
also 
the $\eta$ and the $f_0$ resonances at $M_h \sim 350$ MeV and $M_h \sim 980$ MeV, 
respectively, have been added.
As an example, here 
we illustrate the $z$ and $M_{h}$ dependence of just $D_{1,\rho}^u$:
\begin{align}
     D_{1,\rho}^u(z,M_h; Q_0^2) &= (N_1^{\rho})^2 \, z^{\alpha_1^{\rho}} \, (1-z)^{(\alpha_2^{\rho})^2} \, ( 2 |\mathbf{R}|)^{(\beta_1^{\rho})^2} \notag \\ 
     & \biggl ( \exp \bigl[ - \text{P}(\delta_1^{\rho}, 0, \delta_2^{\rho},0,0;z)\bigr] + (\eta_1^{\rho})^2 \, \text{BW}(m_{\rho} \, , \Gamma_{\rho};M_{h})\biggr ),
     \label{eq:analytic-par}
\end{align}
where $\text{P}(a_1, a_2, a_3, a_4, a_5;z) = \frac{a_1}{z} + a_2 + a_3 \, z\ + a_4 \, z^2 + a_5 \, z^3  $ and  
$ \text{BW}(m,\Gamma ; M_h) = \frac{m^2 \Gamma^2}{(M_h^2-m^2)^2 + m^2\Gamma^2} $. Furthermore, 
isospin symmetry 
is assumed (in particular, $D_1^u =D_1^d $) and
the gluon contribution is taken proportional to the up one, as suggested by single-hadron fragmentation. The total number of parameters is in this case 71.

On the NN side, 
the employed 
architecture 
is 
of type $[2,25,5]$. This corresponds to two input nodes, associated with $z$ and $M_h$, a hidden layer with 25 nodes using a sigmoid activation function, and an output layer with 5 nodes 
(each one for the flavors $q=u,d,s,c$ and one for the gluon, assuming $D_1^q = D_1^{\bar q}$) 
and a quadratic activation function. 
No isospin symmetry is assumed in this case. The total number of parameters is in this case 205.

\section{Unpolarized Results}
We present the results obtained by generating 100 Monte Carlo replicas of the experimental data by fluctuating the data points with Gaussian noise having the same variance as the corresponding experimental uncertainties. As a single estimator of the fit quality, we consider the $\chi^2$ of the mean replica, obtained by averaging over all replicas. For brevity, we mainly discuss the NNLO results, which capture the main features of the analysis~\cite{JHEP2026}.

\begin{figure}[t]
\centering
\includegraphics[width=0.8\textwidth,height=0.32\textheight,keepaspectratio]{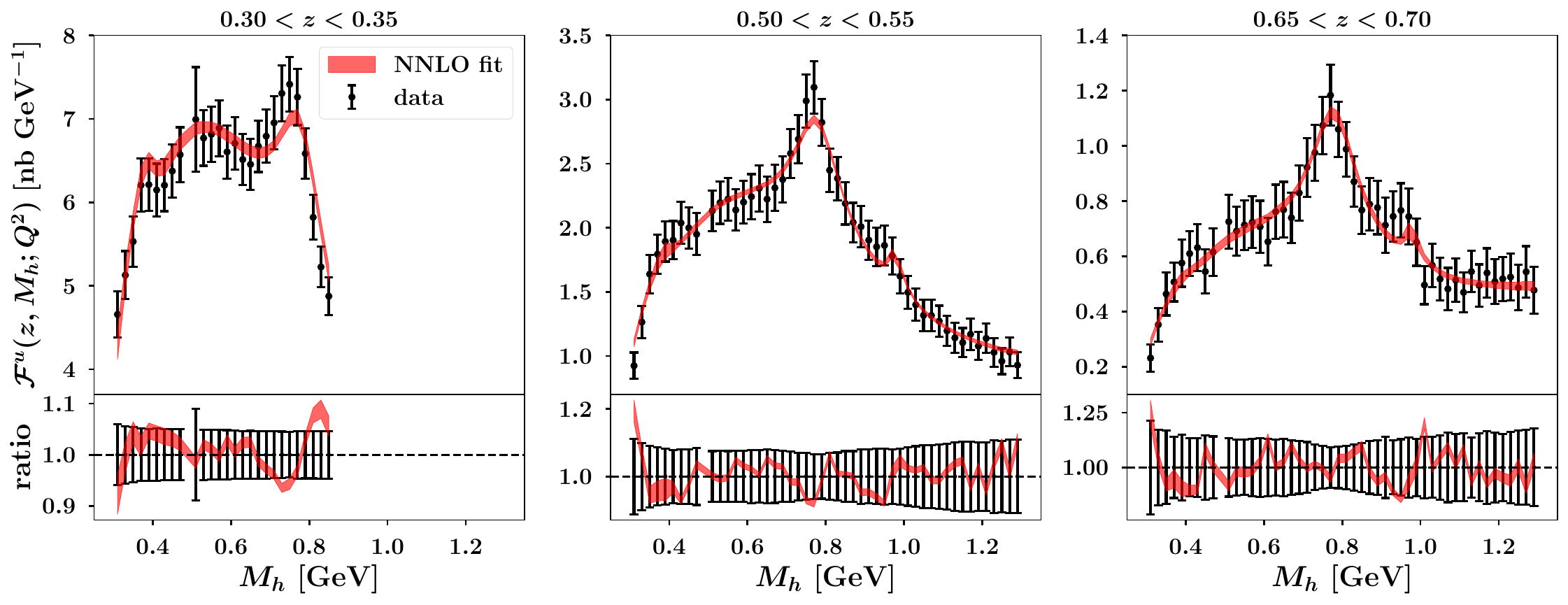}
\vspace{-0.4cm}
\includegraphics[width=0.8\textwidth,height=0.32\textheight,keepaspectratio]{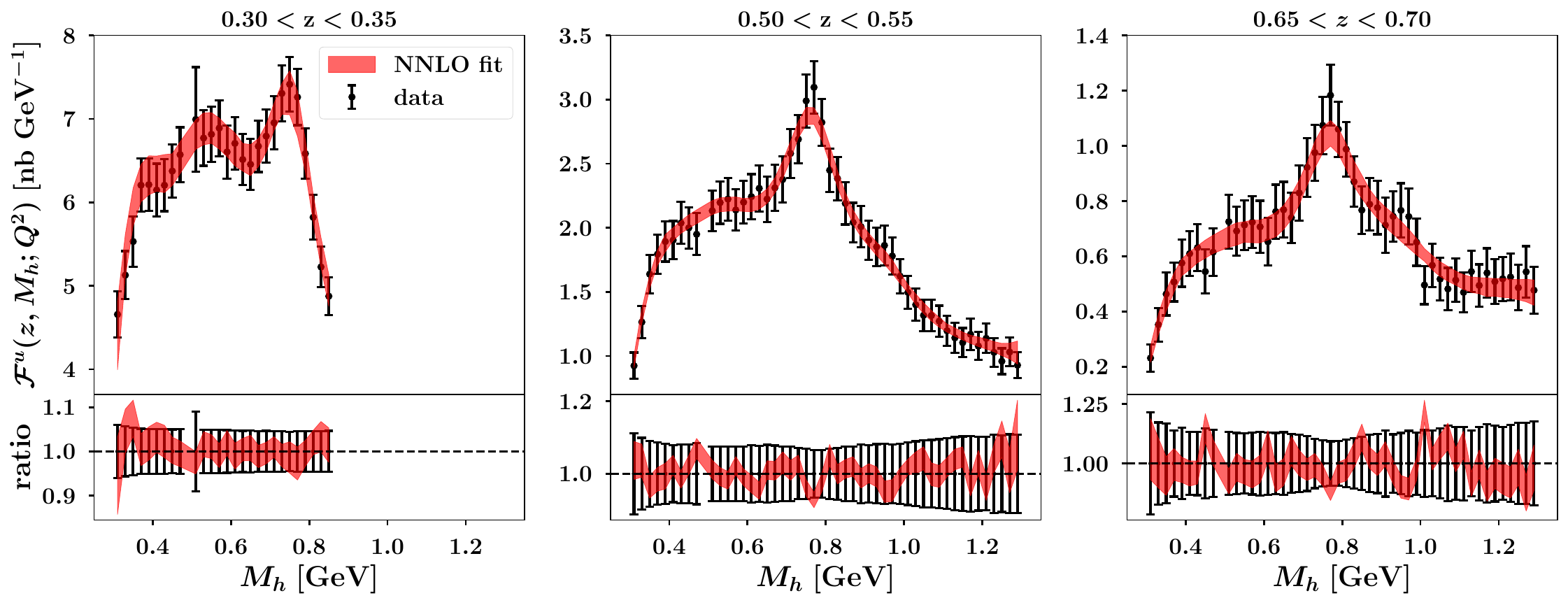}
\caption{Comparison between the NNLO fit obtained with the physics-informed functional form (upper row) and with the NN parametrization (lower row) against the flavor-tagged data of the up pseudodata set, shown as a function of $M_h$ for three different $z$ bins. The uncertainty bands correspond to the 68\% confidence level. For each box, the lower panel shows the ratio between the fit results and the data, normalized to the central value of the latter.}
\label{fig:pred_fixed_orders}
\end{figure}

\begin{figure}[h!]
\begin{center}
\hspace*{-0.05\linewidth}
  \includegraphics[width=0.75\textwidth]{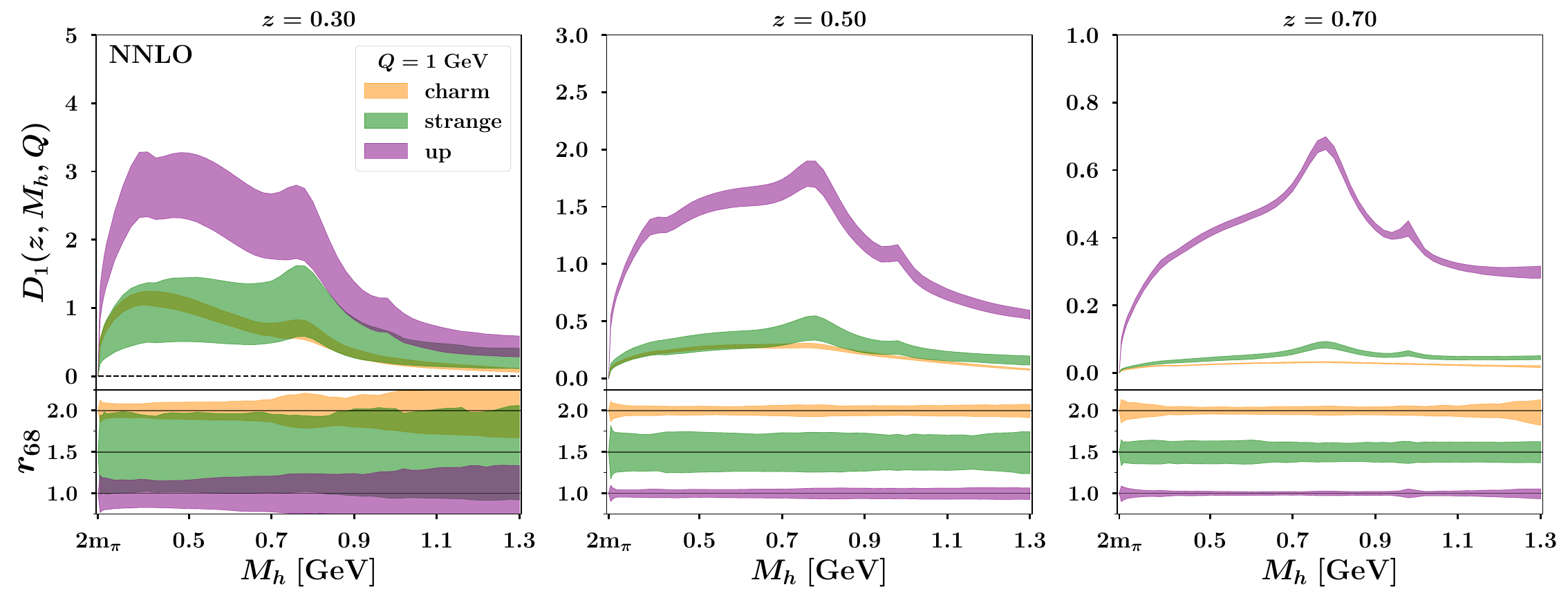}
\hspace*{-0.05\linewidth}
  \includegraphics[width=0.75\textwidth]{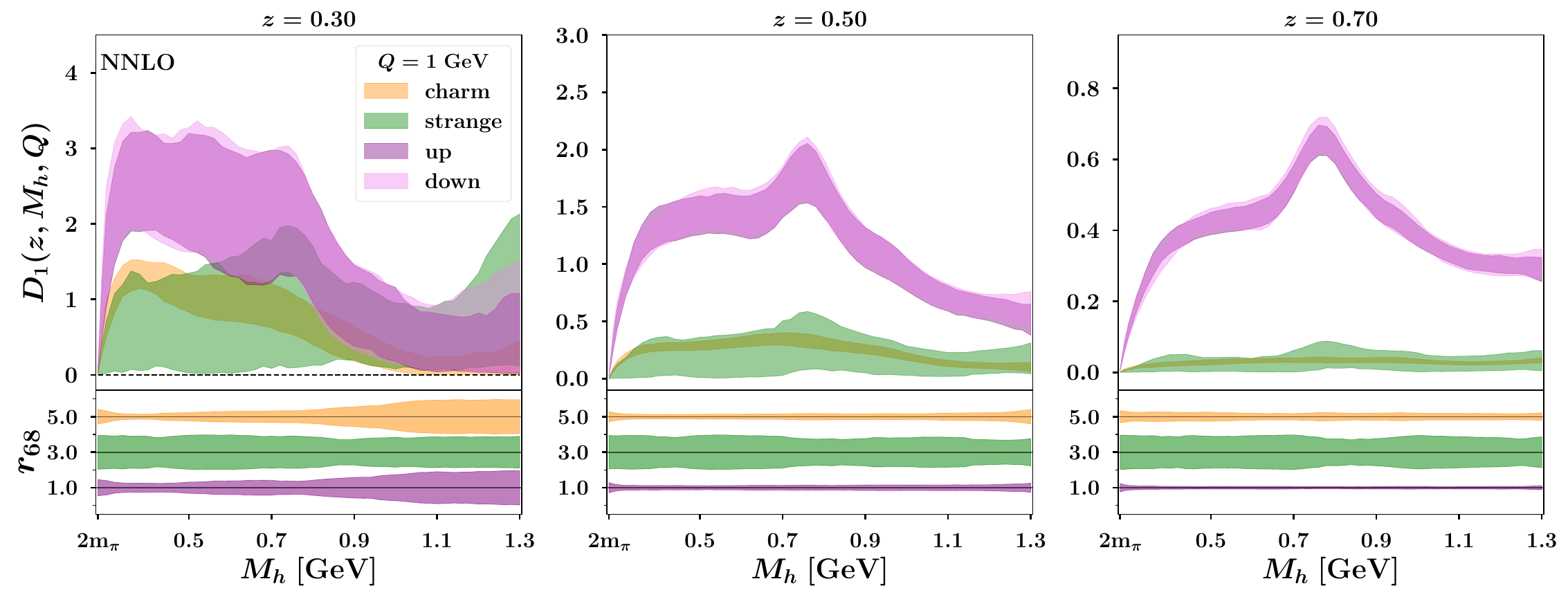}
\end{center}
\vspace{-0.7 cm}
\caption{Comparison between unpolarized DiFFs at NNLO obtained with a physics-informed functional form (upper row) and NN (lower row), as a function of $M_h$ for three different $z$-bins. Uncertainty bands correspond to the 68\% confidence level. For each box, the lower panels show the ratio $r_{68}$ of each replica within the 68 \% band to the central value of the band for $M_h \geq 0.3$. Strange and charm quark results are shifted by 0.5 and 1, respectively.
} 
\label{fig:diff_fixed_orders}
\end{figure}

\begin{figure}[htbp]
    \begin{center}
        \includegraphics[width=0.37\textwidth]{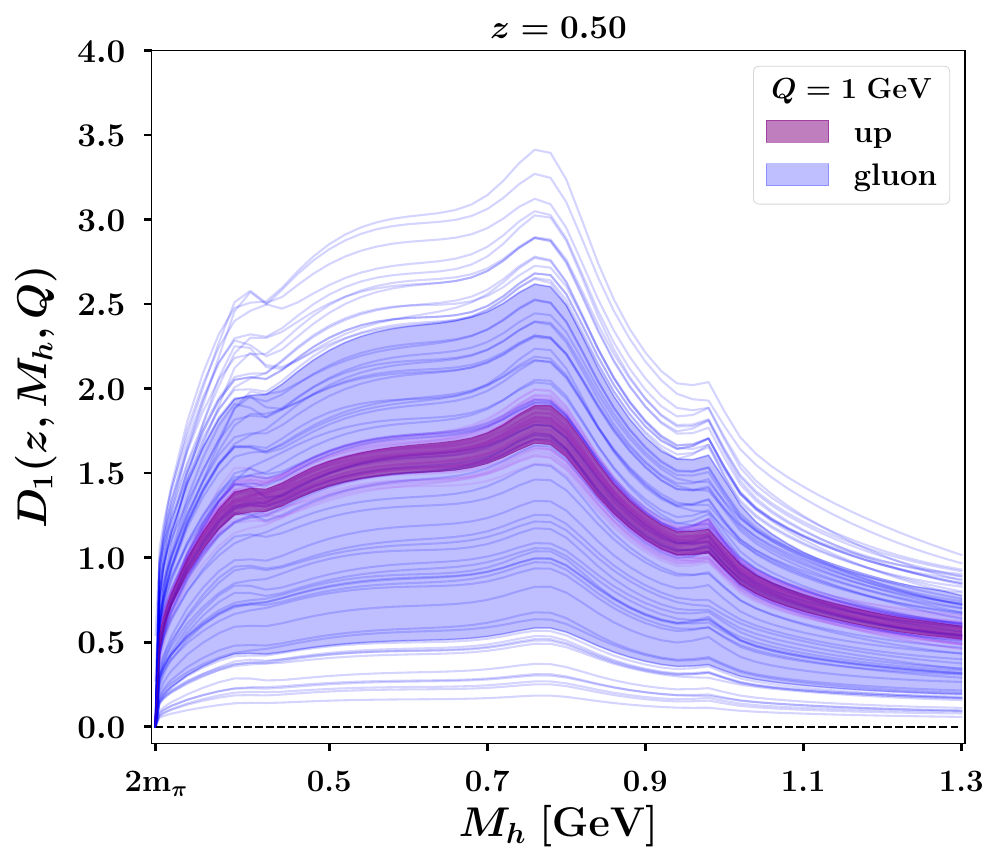}
     \hspace{0.05\textwidth}
        \includegraphics[width=0.34\textwidth]{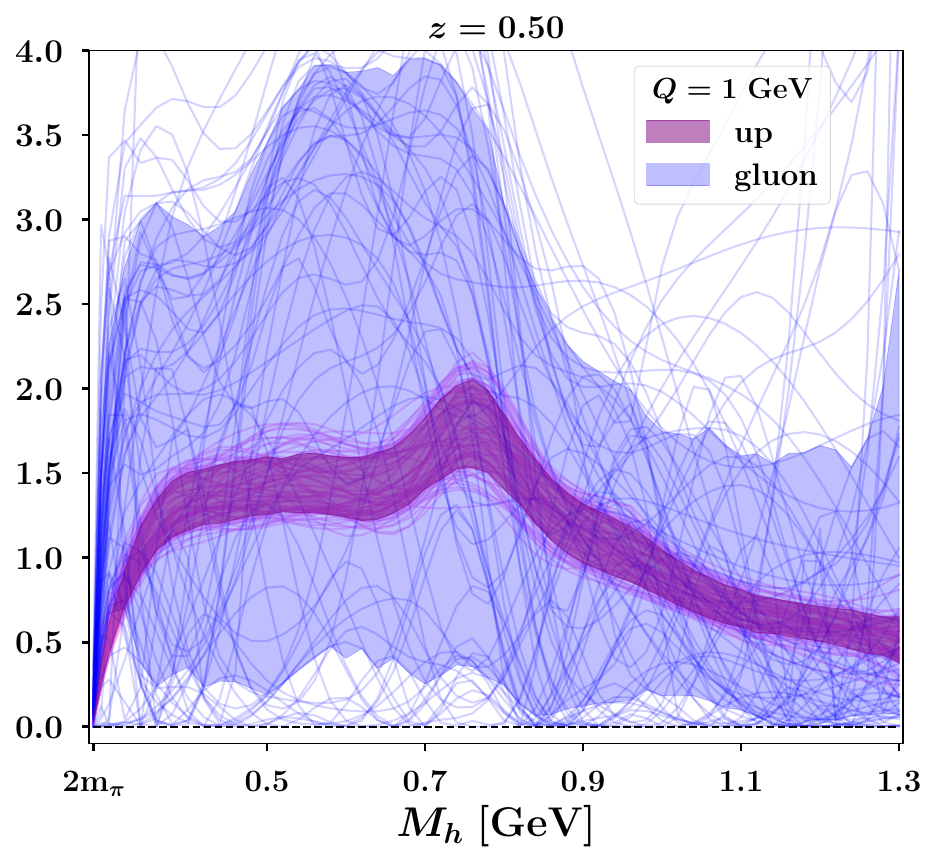}
    \end{center}
\vspace{-0.7 cm}
\caption{The $D_1^i$ with $i=u,g$ at NNLO as function of the pair invariant mass $M_h$ at $Q_0=1$ GeV and $z=0.5$. Uncertainty bands for the 68\% of replicas, with superimposed trajectories of all 100 replicas. Left panel for the fit with the physics-informed functional form, right panel for that with the NN.}
    \label{fig:diff_nnad_allrep}
\end{figure}

For both the physics-informed and NN parametrizations, we obtained  good $\chi^2$ values, namely $\chi^2/N_{\text{data}} = 0.687$ and $\chi^2_{NN}/N_{\text{data}} = 0.535$, respectively. In Fig.~\ref{fig:pred_fixed_orders}, we show the results of the two fits for the up quark. Both approaches show good agreement with the data.
In particular, the NN fit 
seems 
to provide a slightly better description of them, while the physics-informed parametrization reproduces more clearly the resonant structure of the invariant-mass profiles of the cross section, especially in the region of the $f_0$ resonance. The NNLO DiFFs for $q=u,d,s,c$ are shown in Fig.~\ref{fig:diff_fixed_orders}. The uncertainty bands obtained with the two extraction methods are compatible and exhibit the same qualitative behavior. In both approaches, the up and down quark contributions dominate, while the strange and charm quark contributions become negligible at large $z$, since they correspond to unfavored channels for the $(\pi^+\pi^-)$ final state. The NN results for the up and down quarks also support the isospin symmetry assumption imposed in the physics-informed parametrization. As expected from the greater flexibility of the NN parametrization, the NN extraction exhibits larger uncertainty bands, while the resonant structures remain more clearly visible in the physics-informed approach.
The gluon DiFFs at NNLO are shown in Fig.~\ref{fig:diff_nnad_allrep} for  $z=0.5$. In this figure, the gluon uncertainty band is compared with that of the up quark, and the individual replicas are also displayed. In particular, the right panel, corresponding to the NN extraction, shows a very large gluon uncertainty band, allowing for gluon distributions significantly larger than the up-quark contribution. Moreover, the broad spread and erratic behavior of the individual NN replicas indicate that the current data are not sufficient to constrain the gluon contribution. In the physics-informed parametrization, the gluon at the initial scale is chosen to have a smooth shape, similar to that of the up quark, and data do not disprove this assumption.
\section{Polarized DiFFs}
The polarized DiFF $H_1^{\sphericalangle}$ describes the fragmentation of a transversely polarized parton into a hadron pair. As in the unpolarized case, the state-of-the-art extractions are performed at LO, and we are interested in going beyond this accuracy. 
The $H_1^{\sphericalangle}$ can be isolated from the Artru-Collins asymmetry, observed in Double Dihadron Inclusive Annihilation (DDIA) $e^+e^- \rightarrow (\pi^+\pi^-)(\pi^+\pi^-)X$, when two hadron pairs are measured in opposite hemispheres:
\begin{equation}
    A_{e^+e^-} = \frac{\sin^2(\theta)}{1+\cos^2(\theta)} \frac{\sum_q e^2_q \, H_1^{q,\sphericalangle}(z,M_h) \,  H_1^{\bar{q},\sphericalangle}(\bar{z},\bar{M_h})}{\sum_q e^2_q \, D_1^{q}(z,M_h) \, D_1^{\bar{q}}(\bar{z},\bar{M_h})}.
\end{equation}
Therefore, the first step toward the extraction is the computation of the NLO corrections to this observable. In particular, it is important to calculate the partonic cross-section of $e^+e^-\rightarrow q^{\uparrow} \bar{q}^\downarrow$ at NLO. 
We tried to reproduce 
the calculation in Ref.~\cite{CKLM95}, 
but we could not recover their result. 
We therefore decided to perform an independent 
calculation.
At order $\alpha_s$, this requires the evaluation of both virtual and real gluon contributions. The virtual corrections enter through the interference with the LO diagram, while the real contributions arise from gluon emission processes. Summing the two contributions allows one to isolate the collinear singularities, which are then absorbed into the non-perturbative DiFFs through factorization. The calculation was performed in dimensional regularization, and the collinear singularities were subtracted in the $\overline{\text{MS}}$ scheme.
The final result differs from the one reported in Ref.~\cite{CKLM95} and reads:
\begin{align}
\notag
&\dfrac{1}{\sigma_0^{T}} \dfrac{d\sigma}{du\, dz} =  e^2_q \frac{\alpha_s}{2\pi}\, \bigg [ \delta(1-z)P_{qq}^T(u) + \delta(1-u)P_{qq}^T(z) \bigg ]  \ln\biggl( \frac{Q^2}{\mu^2}\biggr)  \, +  e^2_q \frac{4}{3}\frac{\alpha_s}{2\pi} \bigg\{\\
 \notag
   &  
  + \big(\pi^2-8\big) \delta (1-u) \delta (1-z)  \, +  \delta(1-z) \bigg[ (1-u) +2 \, u \, \bigg(\Big(\frac{\log(1-u)}{1-u}\Big)_+ +\frac{\log(u)}{(1-u)_+} \bigg) \bigg]   +\\
   &  +
\delta(1-u) \bigg[ (1-z) +2 \, z \, \bigg(\Big(\frac{\log(1-z)}{1-z}\Big)_+ +\frac{2\log(z)}{(1-z)_+} \bigg) \bigg] \,
 + \frac{2 \, u \, z}{(1-u)_+
   (1-z)_+}+O\big(\epsilon \big) \bigg \} ,
\end{align}
where $u= \frac{P_q \cdot P_{\bar{q}}}{P_q \cdot q_\gamma}$ and $z= \frac{2 P_q \cdot q_\gamma}{q_\gamma^2} $, and $\sigma_0^T$ is the LO transverse-spin coefficient.
The 
first 
term is the 
universal contribution related to the transversity splitting function 
$P_{qq}^T$. 
The second 
scheme-dependent term
corresponds 
to the NLO coefficient 
to 
be implemented 
in a 
NLO 
$H_1^\sphericalangle$ extraction.

\section{Conclusions}
In this talk, we reviewed the results of Ref.~\cite{JHEP2026}, where the extraction of unpolarized DiFFs originally performed in Ref.~\cite{BCRB2012} was revisited using recent BELLE data for $e^+e^- \rightarrow \pi^+\pi^-X$, with the pion pair produced in the same jet. Since the observable is not flavor sensitive, the analysis was supplemented with information from the PYTHIA Monte Carlo event generator.
The study goes beyond LO accuracy for the first time, reaching NNLO precision. Two complementary extractions were carried out in parallel: one with a physics-informed functional form and one with a Neural Network parametrization aimed at reducing theoretical bias. Both methods provide a good description of the data, with small values of $\chi^2/N_{\text{data}}$, and show that up and down quarks dominate over heavier flavors. 
The physics-informed parametrization clearly reproduces the resonant structures of the invariant-mass distributions, which in the Neural Network extraction is not always evident. On the other hand, the Neural Network fit achieves a slightly better description of the data and leads to larger uncertainty bands, consistent with the reduction of theoretical bias.
Both approaches tell us that the gluon DiFF is unconstrained by the current data. This is particularly evident in the Neural Network results, where the uncertainty bands are significantly large, and the individual replicas show erratic behaviors with no clear trend. This highlights the need for data on multiplicities and/or unpolarized cross sections in SIDIS and hadron-hadron collisions.

Finally, we discussed recent developments in the polarized sector of DiFFs. Motivated by inconsistencies found in the NLO calculation of Ref.~\cite{CKLM95} for 
transversely polarized quark-antiquark pair 
production in $e^+e^-$ annihilations, 
we revisited the calculation and obtained a different result. This corrected result is needed to perform for the first time an extraction of $H_1^{\sphericalangle}$ at NLO. Furthermore, we plan to apply the same techniques 
to SIDIS off transversely polarized protons.


\begin{thebibliography}{99}

\bibitem{CRB2007}
F.~A.~Ceccopieri, M.~Radici and A.~Bacchetta,
Phys. Lett. B \textbf{650} (2007), 81-89
doi:10.1016/j.physletb.2007.04.065
[arXiv:hep-ph/0703265 [hep-ph]].

\bibitem{JHEP2026}
V.~Mahaut, L.~Polano \textit{et al.} [MAP (Multi-dimensional Analyses of Partonic distributions)],
JHEP \textbf{02} (2026), 051
doi:10.1007/JHEP02(2026)051
[arXiv:2509.11855 [hep-ph]].

\bibitem{BCRB2012}
A.~Courtoy, A.~Bacchetta, M.~Radici and A.~Bianconi,
Phys. Rev. D \textbf{85} (2012), 114023
doi:10.1103/PhysRevD.85.114023
[arXiv:1202.0323 [hep-ph]].

\bibitem{BCR2013}
A.~Bacchetta, A.~Courtoy and M.~Radici,
JHEP \textbf{03} (2013), 119
doi:10.1007/JHEP03(2013)119
[arXiv:1212.3568 [hep-ph]].

\bibitem{CKLM95}
A.~P.~Contogouris, O.~Korakianitis, Z.~Merebashvili and F.~Lebessis,
Phys. Lett. B \textbf{344} (1995), 370-376
doi:10.1016/0370-2693(94)01319-8


\end{thebibliography}
\end{document}